\documentclass[12pt]{JHEP3}

\usepackage{epsfig}

\newcommand\old{Ref.~\cite{DelDebbio:2002xa}}

\newcommand{\pme}[2]{{}^{+#1}_{-#2}}

\def\MeV{{\rm MeV}}

\def\fm{{\rm fm}}

\def\SUthree{{\rm SU(3)}}

\def\spose#1{\hbox to 0pt{#1\hss}}
\def\ltapprox{\mathrel{\spose{\lower 3pt\hbox{$\mathchar"218$}}
 \raise 2.0pt\hbox{$\mathchar"13C$}}}
\def\gtapprox{\mathrel{\spose{\lower 3pt\hbox{$\mathchar"218$}}
 \raise 2.0pt\hbox{$\mathchar"13E$}}}

\title{
Topological susceptibility from the overlap
}

\author{Luigi Del Debbio \\
	Dipartimento di Fisica dell'Universit\`a 
	di Pisa and I.N.F.N. \\
        Via Buonarroti 2, I-56127 Pisa, Italy \\ 
	E-mail: \email{ldd@df.unipi.it} 
} 

\author{Claudio Pica \\
	Dipartimento di Fisica dell'Universit\`a 
	di Pisa and I.N.F.N. \\
        Via Buonarroti 2, I-56127 Pisa, Italy \\ 
	E-mail: \email{claudio.pica@df.unipi.it} 
} 

\preprint{IFUP-TH 2003/36}

\abstract{ 

The chiral symmetry at finite lattice spacing of Ginsparg-Wilson
fermionic actions constrains the renormalization of the lattice
operators; in particular, the topological susceptibility does not
require any renormalization, when using a fermionic estimator to
define the topological charge. Therefore, the overlap formalism
appears as an appealing candidate to study the continuum limit of the
topological susceptibility while keeping the systematic errors under
theoretical control. We present results for the SU(3) pure gauge
theory using the index of the overlap Dirac operator to study the
topology of the gauge configurations. The topological charge is
obtained from the zero modes of the overlap and using a new algorithm
for the spectral flow analysis. A detailed comparison with cooling
techniques is presented. Particular care is taken in assessing the
systematic errors. Relatively high statistics (500 to 1000 independent
configurations) yield an extrapolated continuum limit with errors that
are comparable with other methods. Our current value from the overlap is
$\chi^{1/4} = 188 \pm 12 \pm 5\, \MeV$.

}


\begin{document}

\section{Introduction}
The QCD lagrangian with $N_f$ massless quarks has a $\mathrm U(N_f)_V
\otimes \mathrm U(N_f)_A$ symmetry at the classical level. The
$\mathrm{SU}(N_f)_A$ subgroup is spontaneously broken, yielding
$N_f^2-1$ massless Nambu-Goldstone bosons with $J^P=0^-$. For $N_f=3$
and small quark masses, the pseudo NG bosons form the light
pseudoscalar mesonic octet observed in the hadronic spectrum. If the
remaining $\mathrm U(1)_A$ subgroup were a symmetry of the theory, one
should observe parity doublets, while the spontaneous
breaking of this symmetry would require the existence of a light
singlet state, the $\eta^\prime$, such that~\cite{Weinberg:ui}:
\begin{equation}
m_{\eta^\prime} \leq \sqrt{3}\, m_\pi.
\end{equation}
The fact that both scenarios are ruled out by the spectrum of light
mesons is known as the $\mathrm U(1)$ problem, whose solution is due
to the anomalous non-conservation of the flavor-singlet axial
current~\cite{'tHooft:up}:
\begin{equation}
\partial_\mu j_\mu^5(x) = 2 N_f\, q(x)
\label{eq:anomaly}
\end{equation}
where $j_\mu^5=\bar\psi(x) \gamma_5 \gamma_\mu \psi(x)$, $q(x)$ is the
topological charge density:
\begin{equation}
q(x) = \frac{g^2}{16 \pi^2} \, \mathrm{Tr }\ F(x) \tilde F(x)
\equiv \partial_\mu K_\mu
\label{eq:topcharge}
\end{equation}
and $K_\mu$ is the current:
\begin{equation}
K_\mu(x) = \frac{g^2}{16 \pi^2} \varepsilon_{\mu\nu\rho\sigma} A_\nu^a(x) \, 
\left(\partial_\rho A_\sigma^a(x) - \frac13 g f^{abc} A_\rho^b(x) 
A_\sigma^c(x) \right).
\label{eq:chern}
\end{equation}

Using the large $N_c$ expansion of the theory~\cite{'tHooft:1973jz},
where $N_c$ is the number of colors, the $\eta^\prime$ mass is related
to the topological susceptibility $\chi$, through the Witten-Veneziano
(WV) formula~\cite{Witten:1979vv,Veneziano:1979ec}:
\begin{equation}
m^2_{\eta^\prime} = \frac{2 N_f}{f^2_\pi} \chi
\label{eq:WV}
\end{equation}
where $f_\pi$ is the pion decay constant, and
\begin{equation}
\chi = \int d^4x\: \left. 
\langle {\mathrm T}\ q(x) q(0) \rangle \right|_\mathrm{YM}
\label{eq:chidef}
\end{equation}
is meant to be computed in the pure Yang-Mills theory. At this stage,
the definition in Eq.~(\ref{eq:chidef}) is still rather abstract,
since a prescription to define the correlator as $x\to 0$ is
necessary. However, Eq.~(\ref{eq:WV}) shows already some
characteristic properties: If $\chi$ goes to a constant in the large
$N_c$ limit, and using the fact that $f_\pi \sim
\sqrt{N_c}$~\cite{'tHooft:1973jz}, one obtains $m^2_{\eta^\prime} \sim
1/N_c$, consistently with the anomalous breaking being suppressed in
that limit. Moreover, the WV relation connects meson spectroscopy to
the nontrivial topological properties of Yang-Mills fields; therefore
it provides a stringent test of the nonperturbative dynamics of QCD,
which can be performed by computing the topological susceptibility
using numerical simulations of the theory defined on a lattice. For
the WV formula to hold, the contact term in the definition of the
topological susceptibility should be such that~\cite{Crewther:1978kq}:
\begin{equation}
\chi = \int d^4x\:\partial_\mu\langle K_\mu(x) q(0)\rangle
\label{eq:crew}
\end{equation}

A lattice topological charge density $q_L(x)$ can be defined by
requiring that the continuum definition is recovered in the na\"{\i}ve
continuum limit, where the lattice spacing $a \to 0$. In the quenched
theory, the lattice operator is related to the continuum one through a
multiplicative
renormalization~\cite{Kronfeld:1987zc,Campostrini:1988cy}:
\begin{equation}
q_L(x) = Z(\beta) a^4(\beta) q(x) + \mathcal O(a^6).
\label{eq:topren}
\end{equation}
where both the lattice spacing $a$ and the renormalization constant
$Z$ are functions of the coupling $\beta=2 N_c/g^2$.  The lattice
topological susceptibility becomes:
\begin{equation}
\chi_L = \sum_x \langle q_L(x) q_L(0) \rangle
\label{eq:latchi}
\end{equation}
however, in order to compare with Eq.~(\ref{eq:WV}), the divergence
for $x \to 0$ has to be properly renormalized, which in principle
involves an additive renormalization to take into account the mixing
with lower dimensional operators~\cite{Campostrini:1989dh}:
\begin{equation}
\chi_L = Z^2(\beta) a^4(\beta) \chi + M(\beta)
\label{eq:chiren}
\end{equation}
Different methods have been devised to deal with the renormalization
of the topological susceptibility. Precise numerical results have been
obtained by computing numerically $Z$ and $M$
non-perturbatively~\cite{DiGiacomo:ba,Alles:ij,Alles:1996nm} or by
using a cooling
technique~\cite{Teper:1985gi,Teper:1985ek,deForcrand:1997sq,
DelDebbio:2002xa}. In the first case, assuming that the UV modes
thermalize more quickly than the long-range topological ones, the
renormalizations are determined by performing a few Monte Carlo
updates on given semiclassical configurations. Indeed, the topological
charge autocorrelation time $\tau_Q$ is known to increase faster than
the autocorrelation time for the Gaussian high-momentum modes as the
continuum limit is approached, which should guarantee that the
short-distance contributions are thermalized without altering the
topological content of the lattice gauge field configuration. Such a
behaviour has been confirmed by recent simulations, showing that
$\tau_Q$ actually grows exponentially with the correlation
length~\cite{DelDebbio:2002xa}. Similarly, the cooling technique also
relies on the hypothesis that the topological and UV degrees of
freedom decouple: in this case one wants to ``switch off'' the
high-momentum modes, thereby obtaining $Z \to 1$ and $M \to 0$,
without modifying the topological content of the theory. The method
relies on the hypothesis that, close to the continuum limit, the
topology of a gauge configuration is determined by the structure of
the fields on length scales $\rho$ larger than the cut-off $a$. If
this were indeed the case, one could ``smooth'' the gauge
configurations on distances $d$ such that $a \ll d \ll \rho$,
e.g. by local minimizations of the action. Along the cooling process
two different phenomena can occur. Instanton-antiinstanton
annihilations can take place, leading to a modification of the
topological charge density, but harmless for the total topological
charge. Besides, a shrinking of the instanton is also observed, which
leads to a loss of topological charge. Metastable results are obtained
if indeed $a \ll \rho$. Despite the success of the cooling method in
its various versions, it still introduces a source of systematic error
in the measurements, which is expected to be larger on coarser lattice,
or at finite temperature.

Lattice formulations of fermions with Dirac operators that satisfy the
Ginsparg-Wilson (GW) relation~\cite{Ginsparg:1981bj} have been
introduced in recent
years~\cite{Narayanan:ss,Narayanan:1994gw,Hasenfratz:1998jp}. They all
have a global chiral symmetry~\cite{Luscher:1998pq} and satisfy an
index theorem~\cite{Hasenfratz:1998ri} at finite cutoff, which allows
the topological charge of a field configuration to be computed from
the number of their zero-modes. In particular, the WV formula has been
derived on the lattice: using GW fermions, one obtains a definition of
the lattice topological susceptibility to use in Eq.~(\ref{eq:WV})
which does not need any subtraction~\cite{Giusti:2001xh}. More details
about the computation of the topological susceptibility from the index
of the overlap Dirac operator are presented in Sect.~\ref{sect:gw}.

Studies of the topological structure of the QCD vacuum using a
fermionic definition of the topological charge have already been
performed both at zero and at finite
temperature~\cite{Giusti:2002rx,AliKhan:2001ym,Edwards:1998sh,
Gattringer:2002mr,Hasenfratz:2002rp,DeGrand:2002gm,Edwards:1999zm,
Cundy:2002hv}. In this paper, we investigate the continuum limit of
the topological susceptibility in the overlap formulation of lattice
QCD; using high statistics and a detailed comparison with the cooling
method, we assess some of the systematic errors that appear in the
computation and in the extrapolation. The results of our simulations
are presented and discussed in Sect.~\ref{sec:results}.

Finally, we summarize our results and discuss some outlooks in
Sect.~\ref{sec:concl}.

\section{Lattice formulation}
\label{sect:gw}

\subsection{Witten-Veneziano formula}
The massless overlap Dirac operator can be written as:
\begin{equation}
D = \frac{1}{\bar a} \left[1 + \gamma_5 \epsilon(H_\mathrm{W}(m))\right]
\label{eq:overlap}
\end{equation}
where $H_\mathrm{W}(m)$ is the Hermitian Wilson-Dirac operator
$\gamma_5\,D_\mathrm{W}(-m)$ with mass parameter $-m$, $\epsilon$ is the
sign function, and $\bar a=a/m$. It is easy to check from
Eq.~(\ref{eq:overlap}) that $D$ verifies the GW relation:
\begin{equation}
\gamma_5 D + D \gamma_5 = \bar a\, D \gamma_5 D,
\label{eq:GW}
\end{equation}
In principle, $m$ can be chosen at will in the range $0<m<2$, with
the different overlap operators defined in this way yielding the same
physics in the continuum limit. In practice, both the spectrum of
$H^2_\mathrm{W}(m)$ and the locality of the overlap depend on
$m$~\cite{Hernandez:1998et}. The dependence on $m$ of the topological
susceptibility is studied in detail and presented below.

In deriving the WV relation on the lattice using GW fermions, 
one obtains~\cite{Giusti:2001xh}: 
\begin{equation}
\frac{f^2_\pi}{2 N_f} m^2_{\eta^\prime} = a^{-4} \sum_x \left. \langle 
\frac{\bar a}{2a} \mathrm{Tr}\left[\gamma_5 D(x,x)\right] \,
\frac{\bar a}{2a} \mathrm{Tr}\left[\gamma_5 D(0,0)\right] \rangle 
\right|_\mathrm{YM}
\label{eq:WVlat}
\end{equation}
A comparison with Eqs~(\ref{eq:WV}) and~(\ref{eq:chidef}) yields for
the lattice topological density:
\begin{equation}
q_L(x) = \frac{\bar a}{2a}\, \mathrm{Tr}\left[\gamma_5 D(x,x)\right]
\label{eq:topden}
\end{equation}
Hence, for the topological charge:
\begin{equation}
Q=\sum_x \frac{\bar a}{2a}\, \mathrm{Tr}\left[\gamma_5 D(x,x)\right] 
= n_- - n_+
\label{eq:index}
\end{equation}
where $n_\pm$ are the numbers of zero modes of the overlap Dirac
operator with positive and negative chirality respectively. Using
Eq.~(\ref{eq:index}), we get for the susceptibility on a finite
lattice of volume $V$:
\begin{equation}
\chi_L = \frac{\langle\left(n_- - n_+\right)^2\rangle}{V}.
\label{eq:chiL}
\end{equation}
and $\chi_L=a^4 \chi$, without additive nor multiplicative
renormalizations needed.

\subsection{Index of the overlap}
Let us now briefly summarize the main features of our implementation
of the overlap operator. In order to study topology, we only need to
concentrate on two aspects: the application of the overlap operator to
a generic spinor field $\psi$, and the computation of the index of
$D$, which requires the study of its
zero-modes. Equation~(\ref{eq:overlap}) can be rewritten as:
\begin{equation}
D = \frac{1}{\bar a} \left[1 + D_\mathrm{W} /\sqrt{H^2_\mathrm{W}(m)}\right],
\label{eq:overlap2}
\end{equation}
so that the implementation of the sign function
$\epsilon(H_\mathrm{W}(m))$ is made explicit.  The Neuberger operator
$D$ is invariant under a rescaling of the Hermitian Wilson-Dirac
operator $H_\mathrm{W}(m)$; the normalization used in this paper is
such that the spectrum of $H_\mathrm{W}(m)$ is bounded by 1. The
implementation of the inverse square root of a sparse matrix with a
potentially large condition number is a demanding task and the
interest in the overlap formulation of lattice QCD has triggered
several studies devoted to its implementation, and a ``standard''
technique has been established. Some of the lowest-lying eigenvalues
of $H_\mathrm{W}(m)$ are computed exactly, and the inverse square root
is trivially defined in the basis provided by the corresponding
eigenvectors. In the orthogonal subspace, $1/\sqrt{H^2_\mathrm{W}(m)}$
is computed either using a polynomial expansion or a rational
approximation. Detailed discussions and an exhaustive list of
references can be found in
Refs.~\cite{vandenEshof:2002ms,Giusti:2002sm}.

In this work, the 15 lowest eigenvalues of $H^2_\mathrm{W}(m)$ are
computed using an accelerated conjugate gradient
algorithm~\cite{Kalkreuter:1995mm} and the corresponding low-lying
modes are treated exactly. The eigenvalues of $H_W^2(m)$ are required
to have an absolute precision of at least $10^{-15}$. The conjugate
gradient search stops when all the eigenvalues have the desired
precision. The error on the eigenvalues is estimated exploiting the
quadratic convergence of the conjugate gradient algorithm in a way
similar to that of Ref.~\cite{Kalkreuter:1995mm}. Since the quadratic
convergence regime is only reached asimptotically, we used the
rigorous bound $|g|$ as an estimate of the error as long as it stays
bigger than $10^{-4}$. Then the error is assumed to be
\begin{equation}
K\frac{\left|\mu-\mu^\prime\right|}
{1-\left[\frac{\left|g\right|}{\left|g^\prime\right|}\right]^2}
\end{equation}
where primed variables refers to previous step values, $\mu$ is the
Ritz functional value and $g$ its gradient.  The numerical constant
factor $K$ was fixed from preliminary study comparing the true error
and the error estimate. The numerical value of $K$ is $10$.  The
\textit{a priori} bound $|g|$ results to be about a factor $1000$
greater than the above estimate.

The approximation on the orthogonal subspace is done using Chebyshev
polynomials, so that the inverse square root reads:
\begin{equation}
1/\sqrt{H^2_\mathrm{W}(m)} = \sum_i \frac{1}{\sqrt{\lambda_i}} |v_i\rangle
\langle v_i| + \sum_n c_n T_n(H^2_\mathrm{W}(m))
\label{eq:approx}
\end{equation}
where $\lambda_i, v_i$ are the eigenvalues and eigenvectors that have
been computed exactly, and $T_n$ are Chebyshev
polynomials. Equation~(\ref{eq:approx}) summarizes the algorithm that
evaluates $\epsilon(H_\mathrm{W}(m)) \psi$, and hence $D \psi$. The
gauge covariance, $\gamma_5$-hermiticity, and locality of our overlap
operator have been succesfully tested~\cite{Hernandez:1998et}.

In order to compute the index of the overlap operator, one has to
compute the eigenvalues of $D^\dagger D$. Since $\left[\gamma_5,
D^\dagger D\right]=0$, one can work in subspaces of given
chirality. Let us denote by $P_\pm$ the projectors on the chirality
subspaces, using the GW relation one can show:
\begin{equation}
D^\dagger D \, P_\pm \equiv P_\pm\, D^\dagger D\, P_\pm  = \frac{2}{\bar
  a} \, D_\pm
\label{eq:subsp}
\end{equation}
where $D_\pm = P_\pm D P_\pm$. Hence, by working in fixed chirality
subspaces, the computation of $(D^\dagger D) \psi$ only requires one
application of the sign function. Moreover, one can show that the
spectra of $D_\pm$ exactly coincide, except for the number of zero
modes in the two sectors. Actually, configurations with zero modes in
both sectors are statistically irrelevant; excluding such a
possibility simplifies the numerical task. Following the strategy
outlined in \cite{Giusti:2002sm}, we run the minimization program in
both sectors simultaneously, until one of them is identified as a
sector in which there are no zero modes.  We require the eigenvalue
$\mu$ to satisfy the bound $\mu - |g|>0$, and the relative precision
of the eigenvalue to be at least $5\%$. A refined search is then
performed in the other chirality sector only, in order to count the
zero modes. A standard conjugate gradient algorithm is used to find
the smallest eigenvalue. This procedure is repeated until an
eigenvalue compatible with that in the other sector is found.  We use
this value as an upper bound for finding the zero modes. A zero mode
is identified with an eigenvalue incompatible within its error with
this upper bound and whose magnitude is less than $20\%$ of the upper
bound. The search for zero mode ends when the current eigenvalue
differs from the upper bound by less than $10\%$ and its relative
precision is at least $10\%$.

\subsection{Spectral flow}
If $D$ is defined using $H_\mathrm{W}(m)$ for some given value of
$m$, then one can show that its index corresponds to the number of
eigenvalues of $H_\mathrm{W}(\mu)$ that cross zero (level crossings),
for $0 \leq \mu \leq m$~\cite{Narayanan:1994gw}. The spectrum of
$H_\mathrm{W}(\mu)$ is characterized by three regions: for small
$\mu$, $0 \leq \mu \leq m_1$, the Wilson-Dirac operator describes
fermions with positive physical mass and hence the spectrum does not
show any level crossing; as $\mu$ is increased a second region $m_1
\leq \mu \leq m_2$ exists, where the spectrum is gapless and crossings
occur; finally the gap should open again for $\mu \geq
m_2$~\cite{Edwards:1998sh}. One would expect that $m_1,m_2 \to 0$, as
the continuum limit is approached. For the values of $\beta$ typically
used for numerical simulations, the gap closes at some value of $\mu
\simeq 0.8$ and remains closed until the end of the allowed region
$\mu=2$. It was pointed out in~\cite{Edwards:1998sh}, that the zero
modes that appear far from $m_1$ have a size of the order of the
lattice spacing and should not affect the physics in the continuum
limit.

To determine the number of level crossing in the spectrum of the
hermitian Wilson-Dirac operator as a function of the mass parameter
$m$ we performed a spectral flow analysis. Using the information about
the spectrum of $H_\mathrm{W}(m)$~\cite{Neuberger:1999pz}, we
introduce a new procedure to locate the crossings in a given mass
region $[m_\alpha, m_\omega]$. Starting with $m=m_\alpha$, we find the
two smallest (in modulus) eigenvalues of $H_\mathrm{W}(m)$ together with their
derivatives with respect to $m$; the value of the mass $m$ is then
increased by $\delta m$, and two new eigenvalues and derivatives are
computed. The algorithm to identify a level crossing in the interval
$\left[m^\prime,m\right]$ (primed variables indicate previous step
values), is the following:
\begin{enumerate}
\item find the two smallest eigenvalues ($\lambda_0<\lambda_1$) and
eigenvectors of $H_\mathrm{W}(m)$;
\item calculate and store eigenvalues derivatives, together with
respective eigenvalues;
\item find which of the two eigenvalues computed in step (1),
$\lambda_0$ or $\lambda_1$, can most probably be considered the
continuation of $\lambda^\prime_0$. This is done by comparing the
average of the derivatives with the incremental ratio of the
eigenvalues;
\item a level crossing between $[m^\prime,m]$ is identified if the
eigenvalue defined in step (3) has a different sign from
$\lambda^\prime_0$ and the eigenvalue derivatives match within
$10\deg$; the exact crossing value and derivative are estimated by
linear interpolation;
\item increment the mass by $\delta m =|\lambda_1|$ and if $m<m_\omega$ repeat
  from step (1)
\end{enumerate}
The eigenvalues at step (1) are found using the same accelerated
conjugate gradient algorithm cited before with a numerical precision
of $10^{-12}$.  If at point (4) there is the possibility of a level
crossing but the derivatives do not satisfy the desidered constraint
we tag the interval and repeat the procedure on that interval with a
smaller mass step size. The value of $\delta m$ in step (5) is chosen
so that only the lowest eigenvalue should possibly have crossed before
$m^\prime$, as granted by the bound on derivatives $|d\lambda / dm| <
1$~\cite{Neuberger:1999pz}.

\section{Numerical results}
\label{sec:results}

\subsection{MC data}
In order to study the continuum limit of the topological
susceptibility, we performed simulations of the \SUthree\ lattice gauge
theory in the standard Wilson formulation, at three values of the
coupling $\beta=5.9,6.0,6.1$. The lattice volume are $12^4,12^4,16^4$
respectively, so that each lattice has a linear size $L\gtapprox 1\,
\fm$. The lattice spacing can be set either from the string tension or
from the low-energy reference scale $r_0$~\cite{Guagnelli:1998ud}. The
relevant numbers for our simulations are summarized in
Tab.~\ref{tab:latspac}, where the dimensionless ratio $r_0
\sqrt{\sigma}$ is also reported. The latter only varies by a few
percent over the range of $\beta$ that we explore, indicating that the
two scales yield compatible results in the continuum limit. Both
quantities will be used to build scaling ratios and study the
continuum limit of $\chi_L$.

For each value of $\beta$, we compute the topological susceptibility
using the index of the overlap, and we study its dependence on the
parameter $m$ by a spectral flow analysis. Furthermore, we also use a
cooling algorithm to evaluate the topological charge on the same
configurations and compare the outcomes. Since the lattice artifacts
in the two definitions can be completely different, there is no reason
a priori for these two quantities to agree at finite lattice
spacing. However, one would like to check that the two methods do
agree as the continuum limit is approached. Moreover, the ``systematic
errors'' of the cooling technique being relatively well understood,
such a comparison, even at finite values of $\beta$, can shed some
light on the effectiveness of the fermionic method in detecting
topology. 

\TABLE[ht]{
\caption{ Lattice spacing in units of the string tension and of the
reference scale $r_0$ for the values of $\beta$ used in this work. The
physical values $a_\sigma$ and $a_{r_0}$ are computed assuming
$\sqrt{\sigma}=440\, \mathrm{MeV}$ and $r_0=0.5\, \mathrm{fm}$
respectively.}
\label{tab:latspac}
\begin{tabular}{llllll}
\hline
\multicolumn{1}{c}{$\beta$}&
\multicolumn{1}{c}{$a \sqrt{\sigma}$}&
\multicolumn{1}{c}{$r_0/a$}&
\multicolumn{1}{c}{$r_0 \sqrt{\sigma}$}&
\multicolumn{1}{c}{$a_\sigma {\mathrm (\fm)}$}&
\multicolumn{1}{c}{$a_{r_0} {\mathrm (\fm)}$}\\
\hline
5.9 & 0.2605(14) & 4.48 & 1.16704 & 0.12 & 0.11 \\ 
6.0 & 0.2197(12) & 5.37 & 1.17979 & 0.10 & 0.09 \\
6.1 & 0.1876(12) & 6.32 & 1.18563 & 0.09 & 0.08 \\
\hline
\end{tabular}
}

The gauge configurations are generated by a Cabibbo-Marinari
algorithm, alternating heatbath and microcanonical updates in a ratio
of 1:4. In order to guarantee the statistical independence of our
configurations, measurements are separated by 1000 updates, which is
much larger than the estimated autocorrelation time, $\tau_Q \ltapprox
200$, at these values of the coupling~\cite{DelDebbio:2002xa}. On each
configuration the topological charge is computed using cooling, a
spectral flow analysis, and the index of the overlap. The number of
configurations analyzed for each lattice is reported in
Tab.~\ref{tab:res1}, together with the values of the topological
charge and susceptibility. It is clear from the raw lattice data, that
the average topological charge is always zero within errors, as it
should be. Moreover, as we shall see in more detail in what follows,
the differences between the different methods do decrease as the
continuum limit is approached. The simulations have been performed on
a cluster of Pentium-4 Xeon processors at 2.2 GHz, using SSE2
instructions to implement the most time-consuming
operations~\cite{Luscher:2001tx}. The total time used for the actual
simulations is about 25 CPU-months; with a breakdown of 15 CPU-months
for the overlap and 10 for the spectral flow. 

\TABLE[ht]{
\caption{
Topological charge $Q$ and susceptibility $\chi_L$.  
}
\label{tab:res1}
\begin{tabular}{llllllll}
\hline
\multicolumn{1}{c}{$\beta$}&
\multicolumn{1}{c}{lattice}&
\multicolumn{1}{c}{$N_\mathrm{conf}$}&
\multicolumn{1}{c}{method}&
\multicolumn{1}{c}{$m$}&
\multicolumn{1}{c}{$Q$}&
\multicolumn{1}{c}{$|Q|$}&
\multicolumn{1}{c}{$\chi~~(\times 10^{4})$}
\\
\hline
5.9 & $12^4$ & 431 & overlap & 1.0 & 0.05(7) & 1.21(5) & 1.27(9)  \\
    &        & 510 & overlap & 1.5 & 0.17(9) & 1.51(5) & 1.83(11) \\
    &        & 432 & cooling &     & 0.01(8) & 1.34(5) & 1.43(9)  \\
    &        &     & \old    &     &         &         & 1.544(7) \\
6.0 & $12^4$ & 956 & overlap & 1.0 & 0.03(4) & 0.90(3) & 0.70(3)  \\
    &        & 956 & overlap & 1.5 & 0.01(4) & 0.98(3) & 0.80(5)  \\
    &        & 956 & cooling &     & 0.03(4) & 0.97(4) & 0.72(3)  \\
    &        &     & \old    &     &         &         & 0.728(5) \\
6.1 & $16^4$ & 440 & overlap & 1.0 & 0.09(8) & 1.12(4) & 0.33(2)  \\ 
    &        & 440 & overlap & 1.5 & 0.09(8) & 1.15(5) & 0.35(3)  \\
    &        & 355 & cooling &     & 0.09(8) & 1.13(5) & 0.34(5)  \\
    &        &     & \old    &     &         &         & 0.382(6) \\
\hline
\end{tabular}
}

\FIGURE[ht]{
\epsfig{file=top-5.9.eps, width=10truecm} 
\caption{
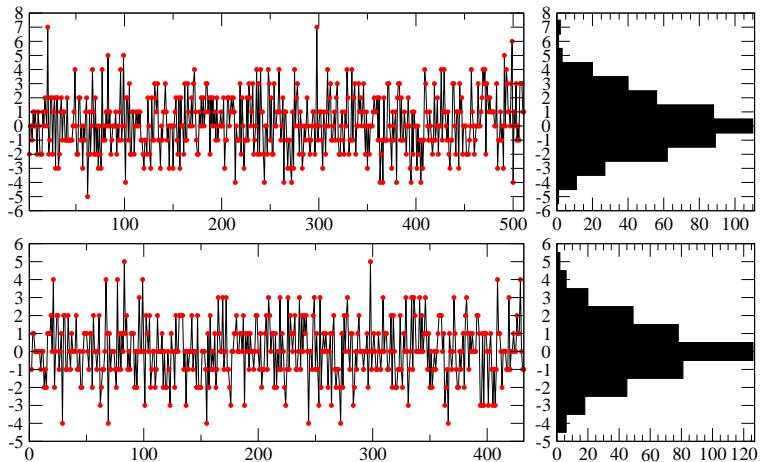
The time history of $Q$ for $\beta=5.9$, m=1 (bottom), and m=1.5 (top).
}
\label{fig:timehist1}
}
\FIGURE[ht]{
\epsfig{file=top-6.0.eps, width=10truecm} 
\caption{
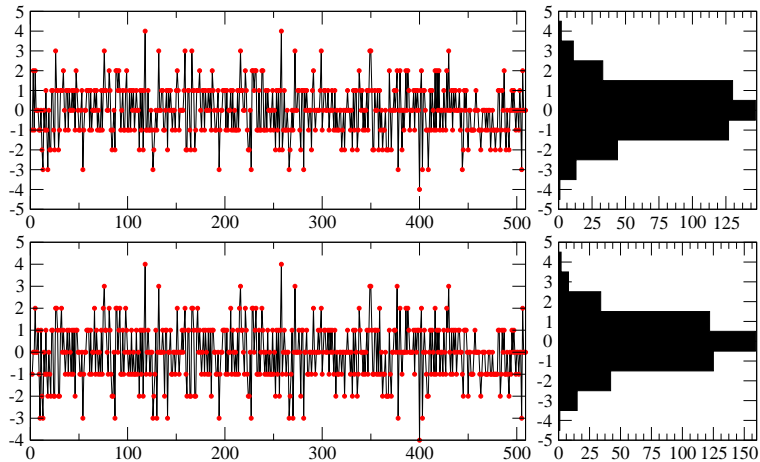
The time history of $Q$ for $\beta=6.0$, m=1 (bottom), and m=1.5 (top).
}
\label{fig:timehist2}
}
The time-history of the topological charge for several values of
$\beta$ and $m$ are shown in Fig.~\ref{fig:timehist1},
\ref{fig:timehist2}, 
where one can see that the algorithm does tunnel from one topological
sector to the other and that the distribution of values is symmetric
and peaked around zero. It is also clear from the figures that the
separation between the measurements is large enough to provide
statistically independent data. At larger $\beta$, the topological
charge distributions are narrower, as one would expect from
semi-classical arguments. Comparing the histograms at different values
of $m$, we see that, as $m$ is increased, the charge distribution
broadens.

The mass dependence of the topological susceptibility, obtained from a
spectral flow analysis, is shown in Fig.~\ref{fig:massdep}.
\FIGURE[ht]{
\epsfig{file=topsusc2.eps, width=10truecm} 
\caption{
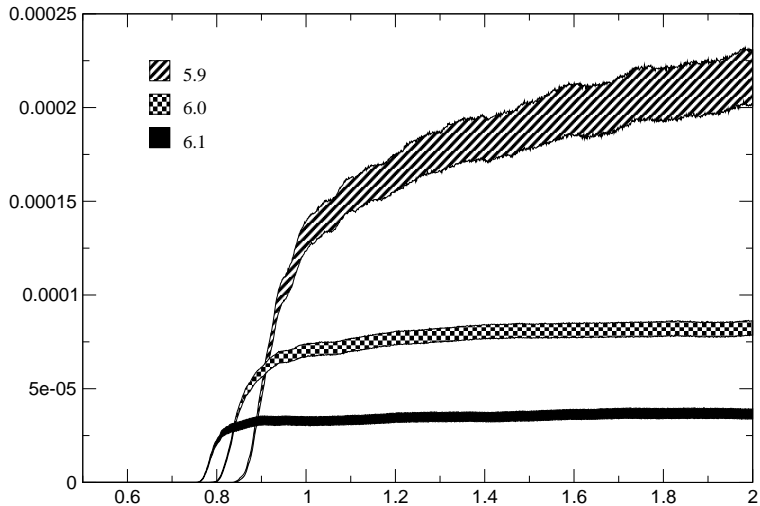
Topological susceptibility $\chi_L$ as a function of $m$ for $\beta=5.9$
(top), $6.0$ (middle), and $6.1$ (bottom).
}
\label{fig:massdep}
}
We observe a large effect at lower values of $\beta$, where the value
of topological susceptibility as a function of $m$ shows a plateau
within the statistical errors only for $m \gtapprox 1.5$. For $m\simeq
1.0$, the value of the susceptibility is about 75\% of the asymptotic
result. For $\beta \geq 6.0$, the same plateau starts for $m\ltapprox
1.0$. Such dependence should be taken into account in studies
performed at fixed value of $m$. The density of crossings vs. $m$ is
reported in Fig.~\ref{fig:crossdensity}; the expected behaviour is
clearly observed. The value of $m_1$, where the gap closes, decreases
as the continuum limit is approached; in the same limit, the level
crossings cluster in the vicinity of $m_1$, so that the topological
susceptibility becomes effectively $m$-independent. For lattice sizes
$L\gtapprox 0.9 \fm$, our results compare well with those presented in
Ref.~\cite{Edwards:1998sh}, where finite size effects were observed on
smaller lattices, e.g. on a $8^3 \times 16$ at $\beta=6.0$.
\FIGURE[ht]{
\epsfig{file=hist.eps, width=10truecm} 
\caption{
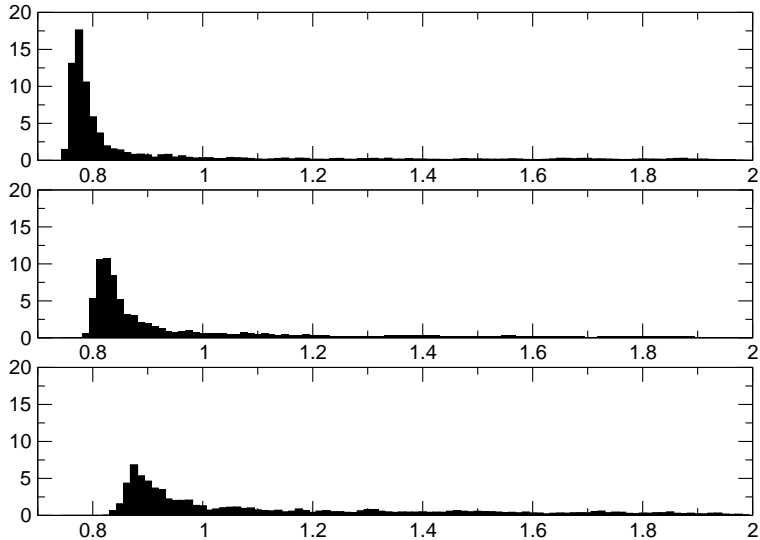
Distribution of level crossings as a function of $m$ for $\beta=5.9$
(bottom), $6.0$ (middle), and $6.1$ (top).
}
\label{fig:crossdensity}
}

The data obtained from the overlap can be compared with those obtained
by a cooling algorithm on the same configurations. Data in
Tab.~\ref{tab:res1} show that the discrepancy between the fermionic
and bosonic estimator of the topological charge ($Q_f$ and $Q_g$,
respectively) decreases as the continuum limit is approached. The
distribution of $\Delta Q=Q_g-Q_f$ is reported in
Fig.~\ref{fig:topvscool} for $m=1$ and $\beta=5.9$ and $6.0$. For
$\beta=6.0$, the agreement of the two methods at $m=1$ is around 90\%;
a comparison of the two values of $\beta$ shows that the discrepancy
between the two methods decreases as the continuum limit is
approached, consistently with the scenario outlined in
Ref.~\cite{Cundy:2002hv}. The agreement between the two results is
also reassuring as far as finite size effects are concerned. Indeed,
the high statistics results obtained in Ref.~\cite{DelDebbio:2002xa}
with a cooling technique show that the finite size effects on the
topological susceptibility at the values of $\beta$ considered in this
work are of the order of a few percent. If the agreement between the
two methods is not a coincidence, the cooling and the fermionic
estimators are expected to display similar finite size effects, which
are known to be well below the statistical accuracy achieved in this
work.
\FIGURE[ht]{
\epsfig{file=deltaQ.eps, width=12truecm} 
\caption{
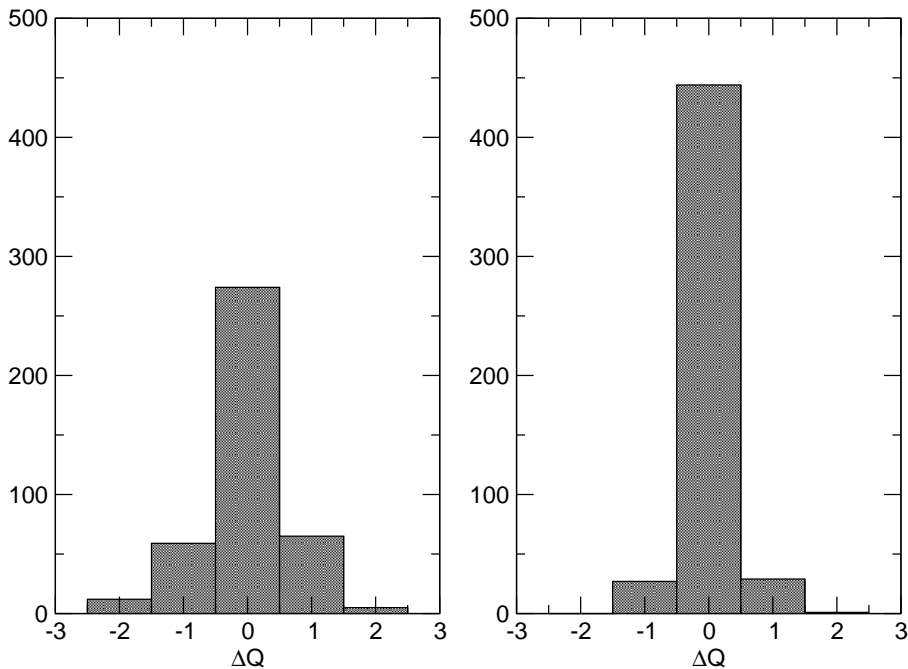
Distribution of the difference $\Delta Q=Q_g-Q_f$ for 
$m=1$, $\beta=5.9$ (left), and $6.0$ (right).
}
\label{fig:topvscool}
}

In Fig.~\ref{fig:topvscoolvsm}, we report the rate of discrepancies
between the fermionic and gluonic definitions as a function of $m$,
for the three values of the coupling. It is interesting to note that
the best agreement is obtained at intermediate values of $m$. The
level crossings in the vicinity of the threshold $m_1$ are the ones
that build the topological charge detected by the cooling
method. Level crossings further away from $m_1$ correspond to
topological excitations that are eliminated by the cooling process,
suggesting that they are associated to modes at the scale of the
cut-off. 
\FIGURE[ht]{
\epsfig{file=cool-sf.eps, width=12truecm} 
\caption{
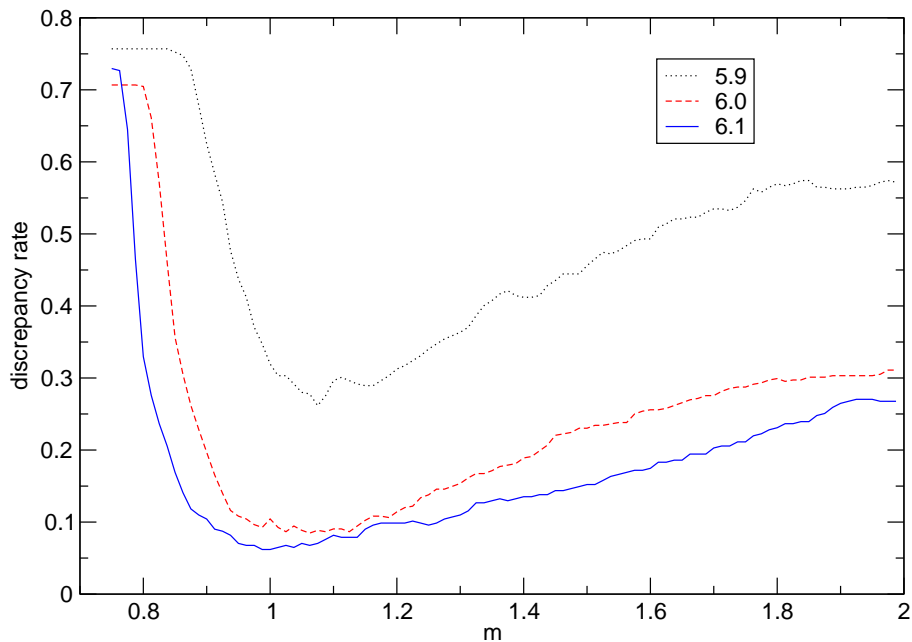
Discrepancy rate between the gluonic and fermionic determinations of
the topological charge as a function of $m$ for $\beta=5.9$ (top),
$6.0$ (middle), and $6.1$ (bottom).
}
\label{fig:topvscoolvsm}
}

\subsection{Extrapolation to the continuum limit}
In order to study the extrapolation to the continuum limit, we build
the adimensional ratios $C_\sigma=\chi_L/\sigma^2$ and $C_{r_0}=\chi_L
r_0^4$, and report their values in Tab.~\ref{tab:res2}. To compare
directly with other works, we also compute the fourth root of the
scaling ratios. The scaling ratio $C_\sigma$ is displayed in
Fig.~\ref{fig:sca1} as a function of $a^2 \sigma$, since the leading
scaling corrections are expected to be $\mathcal O(a^2)$.
\TABLE[ht]{
\caption{
The scaling ratios $C_\sigma$, $C^{1/4}_\sigma$, $C_{r_0}$, and
$C^{1/4}_{r_0}$. 
}
\label{tab:res2}
\begin{tabular}{lllllll}
\hline
\multicolumn{1}{c}{$\beta$}&
\multicolumn{1}{c}{method}&
\multicolumn{1}{c}{$m$}&
\multicolumn{1}{c}{$C_\sigma$}&
\multicolumn{1}{c}{$C^{1/4}_\sigma$}&
\multicolumn{1}{c}{$C_{r_0}$}&
\multicolumn{1}{c}{$C^{1/4}_{r_0}$}\\
\hline
5.9 & overlap & 1   & 0.0273(23) & 0.407(9) & 0.0509(32)& 0.475(8) \\
    & overlap & 1.5 & 0.0390(37) & 0.445(10)& 0.0727(52)& 0.519(8) \\
    & cooling &     & 0.0308(26) & 0.419(9) & 0.0574(36)& 0.489(8) \\
    & \old    &     & 0.0334(9)  & 0.428(3) & 0.0622(4) & 0.499(1) \\
6.0 & overlap & 1   & 0.0300(19) & 0.416(7) & 0.0581(25)& 0.491(5) \\ 
    & overlap & 1.5 & 0.0343(25) & 0.430(8) & 0.0664(33)& 0.508(6) \\
    & cooling &     & 0.0311(19) & 0.420(7) & 0.0602(25)& 0.495(5) \\
    & \old    &     & 0.0312(7)  & 0.420(3) & 0.0604(4) & 0.496(1) \\
6.1 & overlap & 1   & 0.0260(26) & 0.402(10)& 0.0515(38)& 0.476(9) \\ 
    & overlap & 1.5 & 0.0280(28) & 0.409(10)& 0.0555(42)& 0.485(9) \\
    & cooling &     & 0.0282(27) & 0.410(10)& 0.0558(41)& 0.486(8) \\
    & \old    &     & 0.0308(7)  & 0.419(4) & 0.0611(10)& 0.497(2) \\
\hline
\end{tabular}
} 
\FIGURE[ht]{ 
\epsfig{file=scaling.eps,width=12truecm}
\caption{
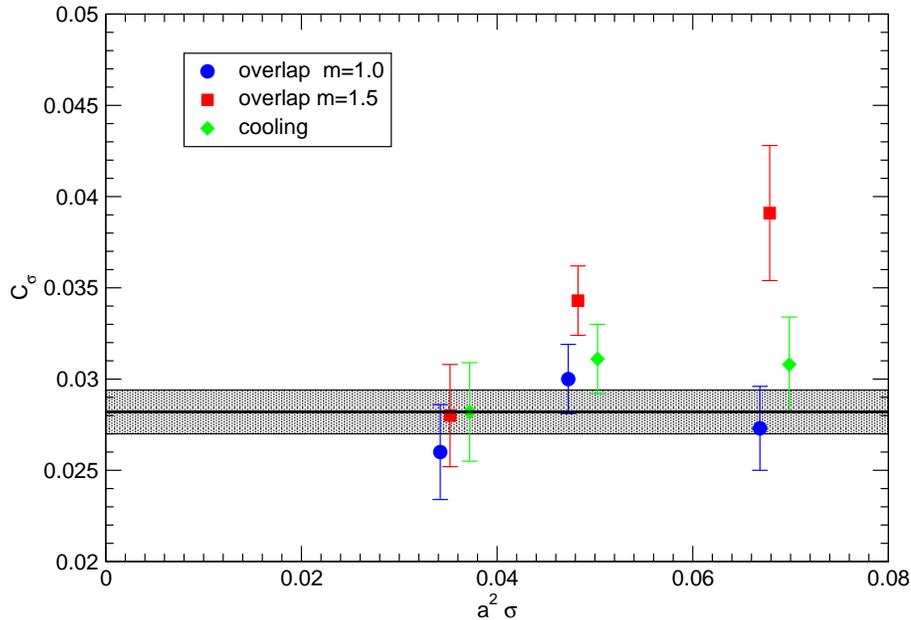
The scaling ratio $C_\sigma$ as a function of $a^2\sigma$. The grey
band is the result of the continuum extrapolation in \old.
}
\label{fig:sca1}
}
The data at $m=1$ do not display any sizeable dependence on $a^2$, and
a naive extrapolation can be obtained simply by quoting an interval
such that all data points are included:
\begin{equation}
C_\sigma(\beta\to\infty) = 0.029 \pm 0.05
\end{equation}
Consistently with the fact that the $m$-dependence becomes negligible
as $\beta$ is increased, as shown in Fig.~\ref{fig:massdep}, the data
point at $m=1.5$ and $\beta=6.1$ is included in the above interval.

Alternatively, one can take a continuum limit by fitting the data
either to a constant, or a linear function. We obtain:
\begin{eqnarray}
C_\sigma(\beta\to\infty) &=& 0.028 \pm 0.02~~~~\mathrm{(constant)} \\
C_\sigma(\beta\to\infty) &=& 0.026 \pm 0.04~~~~\mathrm{(linear)} \\
\label{eq:naiveCL}
\end{eqnarray}
where the errors on the fitted parameters are estimated using a
bootstrap method. The results obtained are in excellent agreement with
previous determinations.

It is clear from Fig.~\ref{fig:sca1} that there are sizeable scaling
violations for $m=1.5$, such that the extrapolated results in the
continuum limit are dominated by the large value at $\beta=5.9$. In
order to get a more reliable result one should increase the statistics
for the current data and add another point closer to the continuum
limit, e.g. at $\beta=6.2$. This would require a larger lattice in
order to avoid finite size effects, an effort that is beyond the scope
of this work.

Data obtained from overlap operators with different values of $m$
can be put together in a single constrained fit, assuming that they
all yield the same continuum limit, and allowing for quartic terms in
the extrapolation:
\begin{equation}
C_{\sigma,m}(\beta) = C_\sigma(\beta\to\infty) + c_{1,m} a^2(\beta) +
c_{2,m} a^4(\beta)
\label{eq:allfit}
\end{equation}
with $c_0$ independent on $m$. Such a procedure allows a non-linear
dependence on $a^2$ to be taken into account; however, one should
remember that the data used for the simultaneous fit are not
independent, so that the bootstrap errors on the fitted values should
only be taken as an indication. The result of the constrained fit is:
\begin{equation}
C_\sigma(\beta\to\infty) = 0.025 \pme{2}{10}
\label{eq:confitres}
\end{equation}
As a consistency check, one can fix $C_\sigma(\beta\to\infty)$ to the
value obtained above, e.g. by fitting the data at $m=1$ to a constant,
and fit the data for both values of $m$ to Eq.~(\ref{eq:allfit}) with
only four parameters. The data fit this ansatz with a
$\chi^2/\mathrm{dof} \approx 1$.

The same analysis can be performed for the scaling ratio
$C_{r_0}$. The reference scale $r_0$ has been computed to great
accuracy and a formula for the interpolation to arbitrary values of
$\beta \in \left[5.7,6.57\right]$ does
exist~\cite{Guagnelli:1998ud}. The use of $r_0$ for the scaling
analysis guarantees a greater uniformity when trying to unify data
from different studies~\cite{Hasenfratz:2002rp}. The results of the
fits are:
\begin{eqnarray}
C_{r_0}(\beta\to\infty)&=&0.054 (3)~~~~~~~\mathrm{for }\ m=1.0,\ 
\mathrm{constant\ fit} \\ 
C_{r_0}(\beta\to\infty) &=&0.055 (8)~~~~~~~\mathrm{for }\ m=1.0,\  
\mathrm{linear\ fit} 
\label{eq:fitr0}
\end{eqnarray}
A naive extrapolation, which includes all the results for $m=1$,
yields $C_{r_0}=0.055 (10)$.

Combining the two determinations, we obtain:
\begin{equation}
\chi^{1/4} = 188 \pm 12 \pm 5\, \MeV
\end{equation}
where the first error is statistical and the second corresponds to the
two different methods of setting the scale.

\section{Conclusions}
\label{sec:concl}
Using a fermionic estimator to compute the topological susceptibility
in the continuum limit has some appealing features: the chiral
symmetry at finite lattice spacing fixes the renormalization of the
lattice operators, so that no renormalization constants are
needed. This is also the case for determinations based on the cooling
technique; nonetheless the latter are based on assumptions on the
separation of gaussian and topological modes.

However, the fermionic method also introduces systematic errors, which
one would like to control: the dependence on the mass parameter $m$,
that appears in the definition of the overlap, finite size effects,
and the size of the lattice artifacts at current values of the
coupling. 

In agreement with theoretical arguments and previous numerical
investigations, we find that the topological suceptibility does vary
with $m$ at small values of $\beta$, and that this dependence
disappears as the continuum limit is approached. Studies using the
overlap at fixed value of $m$ should take this source of systematic
error into account, even though we expect the magnitude of the
systematics to depend on the observables. Finally, comparison with
results obtained using a cooling procedure on the same configurations
shows that the two determinations agree in the continuum limit. For
values of $m\gtapprox 1.2$, the fermionic method detects topological
charges that are smoothed away by the cooling procedure, which could
be related to degrees of freedom at the cut-off scale.

On our lattices of linear size $L\simeq 1 \fm$, the comparison with
cooling suggests that finite size effects are small compared with our
current statistical error. 

Our results for the continuum extrapolation are compatible with other
determinations. In order to further constrain our extrapolation, one
more point at $\beta=6.2$, larger lattices, and higher statistics are
needed. Moreover, improved actions could be used to reduce lattice
artifacts and possibly achieve a faster convergence towards the
continuum limit. Based on our experience, a precise determination of
the continuum limit extrapolation is well within reach.

If one is only interested in the topological charge and
susceptibility, a spectral flow analysis is sufficient and can be
implemented quite efficiently using the procedure outlined
above. Instead studies of the overlap operator are needed if one wants
to investigate the correlator of the topological density $\langle
q_L(x) q_L(0) \rangle$.

\acknowledgments This work is partially supported by INFN and by
MIUR -- Progetto ``Teoria e Fe\-no\-me\-no\-lo\-gia delle Particelle
Elementari''. The authors acknowledge interesting discussions with
Adriano Di Giacomo. LDD is grateful to Martin L\"uscher and Ettore
Vicari for many discussions, and to the Theory Division at CERN for
hospitality and financial support at various stages of this work. Last
but not least, LDD thanks Rovena Medei for her financial support.

\end{document}